\documentclass[aps,pre,twocolumn,showpacs,amsfonts,amssymb,floatfix,superscriptaddress]{revtex4-1}
\usepackage[utf8]{inputenc}
\usepackage{float}
\usepackage{graphicx}
\usepackage{amssymb}
\usepackage{amsmath}
\usepackage{bbm}

\makeatletter

\newcommand*{\diff}{\mathop{\!\mathrm{d}\!}}
\newcount\hour \newcount\minute
\hour=\time \divide \hour by 60 \minute=\time \count99=\hour
\multiply \count99 by -60 \advance\minute by \count99 \hour=\time
\divide \hour by 60
\date{August 7, 2013} 

\makeatother

\begin{document}

\title{Critical manifold of globally coupled overdamped anharmonic oscillators\\
driven by additive Gaussian white noise}

\author{Rüdiger Kürsten}
\email{ruediger.kuersten@itp.uni-leipzig.de}
\affiliation{Institut für Theoretische Physik, Universität Leipzig, POB 100 920, D-04009 Leipzig, Germany}
\affiliation{International Max Planck Research School Mathematics in the Sciences, Inselstraße 22, D-04103 Leipzig, Germany}
\author{Susanne Gütter}
\affiliation{Institut für Theoretische Physik, Universität Leipzig, POB 100 920, D-04009 Leipzig, Germany}
\author{Ulrich Behn}
\email{ulrich.behn@itp.uni-leipzig.de}
\affiliation{Institut für Theoretische Physik, Universität Leipzig, POB 100 920, D-04009 Leipzig, Germany}
\affiliation{International Max Planck Research School Mathematics in the Sciences, Inselstraße 22, D-04103 Leipzig, Germany}

\begin{abstract}
	We prove for an infinite array of globally coupled overdamped anharmonic oscillators subject to additive Gaussian white noise the existence of a well-behaved critical manifold in the parameter space which separates a symmetric phase from a symmetry broken phase. Given two of the system parameters there is an unique critical value of the third. The proof exploits that the critical control parameter $a_c$ is bounded by its limit values for weak and for strong noise. In these limits the mechanism of symmetry breaking differs. For weak noise the distribution is Gaussian and the symmetry is broken as the whole distribution is shifted in either the positive or the negative direction. For strong noise there is a symmetric double-peak distribution and the symmetry is broken as the weights of the peaks become different. We derive an ordinary differential equation whose solution describes the critical manifold. Using a series ansatz to solve this differential equation we determine the critical manifold for weak and for strong noise and compare it to numerical results. We derive analytic expressions for the order parameter and the susceptibility close to the critical manifold.
\end{abstract}
\maketitle

\section{Introduction}

Nonlinear globally coupled systems under the influence of noise have been an active
field of research over the last few decades \cite{SJG07,GS99}. For additive noise there is a natural and far-reaching analogy to
equilibrium thermodynamics \cite{Fr05}.

We consider an array of $L$ harmonically coupled overdamped anharmonic 
oscillators subject to additive noise which is governed by the system
of Langevin equations
\begin{align}
\dot{x_{i}}  = 
ax_{i}-x_{i}^{3}-\frac{D}{L-1}\sum_{j=1}^L\left(x_{i}-x_{j}
\right)+\xi_{i}(t) \label{eq:eqofmotion}
\end{align}
for $i = 1, \cdots , L$.  Each of the oscillators is  harmonically
coupled to all the others (global coupling), the total strength of the
coupling is $D$. The additive noise $\xi_{i}(t)$ is a zero mean
Gaussian white process with autocorrelation
\begin{align}
\langle\xi_{i}\left(t\right)\xi_{j}\left(s\right)\rangle =  \sigma^{2}\delta_{ij}\delta\left(t-s\right),
\end{align}
where $\sigma^{2}$ denotes the noise strength. The steady state of an
isolated system without noise ($D=0, \sigma =0$) undergoes a pitchfork
bifurcation if the control parameter $a$ changes the sign. The isolated system ($D=0$) describes
diffusion in a potential with one or two minima depending on the sign of $a$ which is thoroughly studied since \cite{Kr40}.
In the limit $L \to \infty$ the array exhibits a continuous phase transition
with a critical point $a_c(D, \sigma)$ and the mean field critical exponent $1/2$ such that for $a \le a_c$ the stationary
probability density $p(x)$ is symmetric with respect to $x=0$, and for $a>a_c$
it is non-symmetric such that the order parameter $\langle x \rangle \ne 0$.

The dependence of $a_c$ on the other parameters has been investigated already
by a number of authors
\cite{KS75,DZ78,Da83,BPAH94,GPSB96}.
Kometani and Shimizu \cite{KS75} closed the equation of motion for the
moments using a decoupling which is correct for Gaussian
distributed variables. The critical point is determined by the occurrence of a
non-trivial solution. This yields, in our notation, $a_c=3 \sigma^2/(2D)$
which is asymptotically correct for weak noise, see below.
Desai and Zwanzig \cite{DZ78} described the system by a self-consistent dynamic
mean field theory. They evaluated numerically the correct phase transition
condition and observed that the critical point deviates from the 
result of a Gaussian approximation.
Dawson \cite{Da83} correctly claimed existence and uniqueness of the critical parameter. We show that he used a wrong argument and give a different proof of uniqueness. He proved in the limit of infinitely many oscillators that the fluctuations at the critical point are non-Gaussian and occur at a slower time scale than the noncritical fluctuations. Furthermore, he observed that for $D=a$ the critical
point can be computed up to a quadrature. 
Van den Broeck and collaborators \cite{BPAH94} also gave numerical results
for the parameter dependence of $a_c$ in mean field theory and
compared this with simulations for a system with nearest
neighbor coupling in $d=2$. Implicitly they state that for strong
coupling ($D \rightarrow \infty$) the critical point is $a_{c} =0$, cf. also
\cite{GPSB96}.

Shiino \cite{Sh85,Sh87} proved a $H$-theorem showing thus that the
stationary state is asymptotically reached for long times and analyzed the
stability of the trivial and the symmetry breaking solutions. 

The harmonic coupling between the
constituents of an array was introduced by Kometani and Shimizu
\cite{KS75} to describe the interaction between myosin and actin filaments
in muscle contraction. In this context the variables $x_i$ represent velocities rather than coordinates and the system can be regarded as an early example of a canonical-dissipative system \cite{ES05}, which have been proposed to describe, e.g., swarm dynamics \cite{SET01}.

The harmonic coupling between nearest neighbors of a regular lattice 
can be conceived as discretization of the Laplace operator. Thus, in the continuum limit there is a relation to mean field solutions of a class of models
described by stochastic partial differential equations, see e.g. \cite{HSL07}.

There is further a relation to the discretized version of the $\Phi ^{4}$-Ginzburg-Landau model, in different context also known as
soft-spin Ising model. For example, in \cite{SZ81} the authors studied spin glasses, where the coupling strength for each pair of coordinates $x_i$, $x_j$ is an independent Gaussian distributed random variable.

Noise induced phenomena in a double-well potential are still a topic of recent research. For example, in \cite{HNV12} the authors investigated a Fokker-Planck equation driven by dynamical constraints, modeling many-particle sto\-rage systems.

For a similar system driven by multiplicative noise instead of additive noise also continuous phase transitions occur. The critical exponent of the order parameter $\langle x\rangle$ undergoes a transition from a constant (though non-mean field) value towards a parameter-dependent value when $\sigma^2/(2D)$ exceeds a threshold \cite{BLMKB02}. Also higher moments $\langle x^m \rangle$ show such transitions \cite{MCC05,SRHM07}. It is a natural question whether this behavior is robust against additive noise, since in natural systems additive noise is apparently unavoidable. Although many papers study systems with additive and multiplicative noise (for early refs. see \cite{BPAH94,GPSB96}) this question has not been explicitly addressed. To determine critical exponents it is advantageous to know analytically the parameter dependence of the critical manifold. Besides our general interest this is an additional motivation to study the present system.

The paper is organized as follows. In Sec. \ref{sec:model} we reformulate the
model in  the Fokker-Planck picture and explain the self-consistent mean field
approach which becomes exact in the limit of infinite system size. We show that the
self-consistency condition is equivalent with the stationarity condition for the
center of mass variable.

In Sec. \ref{sec:critmnf} we prove the existence of a well-behaved critical
manifold in the parameter space which separates the regime with $\langle
x\rangle =0 $ from the regime with broken symmetry. The critical manifold is determined by an implicit integral equation, the phase transition condition (PTC). To show its wellbehavedness it is
necessary to know that the critical parameter $a_c(D, \sigma)$ is
bounded by values which are asymptotically reached in the limits of strong and
weak noise, respectively. The wellbehavedness of the critical manifold
allows to reduce the number of parameters by rescaling and implies that
$a_c/D$ is only a function of the ratio $\sigma/D$. Thus it exhibits the same
behavior for strong coupling as for weak noise, and for weak coupling as for
strong noise. 

Exploiting that the critical manifold is well-behaved, together with exact relations of moments on this mani\-fold we derive an ordinary differential equation (ODE) for $a_c/D$ as a function of $(\sigma/D)^2$. To determine a solution for small and large values of $(\sigma/D)^2$ we use a series expansion. The results from the ODE are the same as from an evaluation of the PTC using the Laplace method but easier to obtain. The asymptotic behavior of the series agrees with that of the numerical solution of the PTC. We were not able to prove convergence of these series, but for small $(\sigma/D)^2$ a Pad\'e approximant agrees well with the numerical solution. We also give results for a series expansion around $a=D$ which is analytically treatable.

%
In Sec. \ref{sec:critbeh} we determine the behavior of the order parameter and
the susceptibility near the critical manifold. Thanks to the boundedness of
$a_c$ the existence of a tricritical point is excluded and the critical
exponents are the mean field exponents, as expected. We derive
the amplitudes of the power laws of order parameter and susceptibility in
closed form in terms of $a_c$. The amplitude ratio of the susceptibilities
just above and below the critical point is universal as expected by 
analogy with equilibrium second order phase transitions.

Several detailed calculations and technical discussions are deferred to the
appendices. In Appendix \ref{sec:moments} recursion relations for the moments are
given. In Appendix \ref{sec:proofac} we derive in some detail the bounds of $a_c$. In
Appendix \ref{sec:laplace} the phase transition condition is evaluated by the
Laplace method up to the third leading order in the limits of weak and strong
noise, respectively.

\section{The Fokker-Planck Picture \label{sec:model}}

The Fokker-Planck equation corresponding to Eq.~\eqref{eq:eqofmotion} is
\begin{align}
	\partial_t p(\mathbf{x}, t) = & \sum_{i=1}^{L}-\partial_{x_i}\Big\{ \Big[ (a-D)x_i -x_i^3  + \frac{D}{L-1} \sum_{j=1}^{L}x_j\notag \\
	& - \frac{\sigma^2}{2}\partial_{x_i}\Big] p(\mathbf{x}, t)\Big\}.
	\label{eq:fpldim}
\end{align}
Integrating over all coordinates but $x_1$ we find
\begin{align}
	\partial_{t}p_1^L(x_1, t)  = & -\partial_{x_1}\Big\{ \big[ (a-D)x_1 -x_1^3  + D \langle x_2|x_1\rangle \notag \\
	& - \frac{\sigma^2}{2}\partial_{x_1} \big] p_1^L(x_1, t)\Big\},
	\label{eq:fpstatldim}
\end{align}
where $\langle x_2 | x_1 \rangle $ is the conditional expectation value of $x_2$ given $x_1$ and $p_{1}^{L}(x_1, t)$ is the probability distribution of $x_1$ for the system with $L$ constituents.
For $L \rightarrow \infty$ we assume independence of coordinates, i.e.
\begin{align}
\langle x_2|x_1\rangle \approx \langle x_2\rangle = \langle x_1\rangle
\label{eq:approximation1}
\end{align}
and find the one-particle distribution as the solution of
\begin{align}
\partial_t p(x,t) =  & -\partial_x \Big\{ \Big[ (a-D) x -x^3 + D \int_{-\infty}^{\infty}\diff x'  x' p(x',t) \nonumber \\
 & -\frac{\sigma^2}{2} \partial_x \Big] p(x,t)  \Big\},
	\label{eq:statsol}
\end{align}
where we wrote $p(x,t)$ instead of $p_{1}^{L\rightarrow \infty}(x_1, t)$.

\indent Besides Eq.~\eqref{eq:approximation1} there is a more rigorous approach. Following \cite{Sh87}, we introduce the empirical distribution defined by
\begin{align}
	p^{L}_{e}(x,t)\diff x  = & \frac{1}{L} \sum_{i=1}^{L} \mathbbm{1}_{[x,x+\diff x]}(x_i(t)),
\end{align}
which is an probability-distribution-valued random variable.

\indent Provided that for $t=0$ all coordinates are independent and identically distributed with distribution $p_0(x)$, it is a result of \cite{Da83} that in the limit $L \rightarrow \infty$ both distributions converge weakly to $p(x,t)$, which is the unique solution of Eq.~\eqref{eq:statsol} with initial condition $p(x, t=0) = p_0(x)$. Hence, for large $L$ we can interpret $p(x,t)$ either as the empirical distribution of the many coordinates or as the probability distribution of a single coordinate.

For an infinitely large system the harmonic coupling in Eq. \eqref{eq:eqofmotion} of the site $i$ to all other sites becomes
\begin{align}
	- \frac{D}{L-1} \sum_{j=1}^{L}\left(x_i - x_j\right) \rightarrow -D\left(x_i - m\right),
	\label{eq:interactionterm}
\end{align}
where
\begin{align}
&	m(t) = \lim_{L\rightarrow \infty} \frac{1}{L} \sum_{j=1, j\neq i}^{L} x_j \nonumber \\
 = \lim_{L\rightarrow \infty} &\int_{-\infty}^{\infty}\diff x \,  x p^{L}_{1}(x,t) = \int_{-\infty}^{\infty}\diff x \,  x p(x,t)
\label{eq:selfcons}
\end{align}
is the mean field exerted by all other sites to the system at site $i$.

It was shown in \cite{Sh85,Sh87} that the time dependent solution of Eq. \eqref{eq:statsol} approaches the stationary solution \eqref{eq:statdistr} for long times. In the stationary case $m$ is constant and can be considered as a parameter. We find the stationary solution of Eq.~\eqref{eq:statsol}
\begin{align}
p_s(x;m) = \frac{1}{Z} \exp \left[\frac{2}{\sigma^2}\left( (a-D) \frac{x^2}{2}-\frac{x^4}{4} + D m x \right) \right]\label{eq:statdistr},
\end{align}
with normalization
\begin{align}
Z = \int_{-\infty}^{\infty}  \diff x \exp \left[\frac{2}{\sigma^2}( (a-D)\frac{x^2}{2}-\frac{x^4}{4} + D m x ) \right],
\label{eq:norm}
\end{align}
satisfying
\begin{align}
	p_{s}(x;m) = p_{s}(-x;-m).
	\label{eq:symstat}
\end{align}
The mean field should solve the self-consistency equation
\begin{align}
	m = & \int_{-\infty}^{\infty} \diff x \, x p_s(x; m).
	\label{eq:scc}
\end{align}

Obviously the mean field is related to the center of mass coordinate
\begin{align}
R = & \frac{1}{L} \sum_{i=1}^{L} x_i,
\end{align}
obeying the Langevin equation
\begin{align}
\dot{R}  = & \frac{1}{L}\sum_{i=1}^{L}\left( ax_i -x_i^3 \right) + \frac{1}{L} \sum_{i=1}^{L} \xi_i(t) .\label{eq:langevincom}
\end{align}
For the infinite system the noise term vanishes due to the law of large numbers and Eq. \eqref{eq:langevincom} becomes
\begin{align}
	\lim_{L \rightarrow \infty} \dot{R} = a \langle x\rangle - \langle x^3\rangle.
	\label{eq:langcomlim}
\end{align}
In the stationary case $a \langle x\rangle - \langle x^3\rangle =0 $. This is equivalent to the self-consistency condition \eqref{eq:scc} as can be seen writing
\begin{align}
	& \int_{-\infty}^{\infty} \diff x \, \left(ax - x^3\right)p_{s}(x; m)  \notag \\
	&= \frac{1}{Z} \frac{\sigma^2}{2}\int_{-\infty}^{\infty} \diff x \,  \left\{ \partial_{x} \exp\left[\frac{2}{\sigma^2} \Big(\frac{a}{2}x^2-\frac{1}{4}x^4 \Big) \right] \right\} \notag \\
 & \; \; \; \; \; \times \exp \left[ \frac{2D}{\sigma^2}\Big(-\frac{1}{2}x^2+mx \Big) \right] \notag \\
 & = D\left(\langle x\rangle-m\right) = 0 . \label{eq:limR}
\end{align}
The second equality follows by partial integration and observing that the boundary term vanishes since $p_{s}(x; m) $ decays exponentially fast for $|x|\rightarrow \infty$.

\section{The Critical Manifold\label{sec:critmnf}}
\subsection{Existence and General Properties\label{subsec:cmgeneral}}
\indent We introduce the function
\begin{align}
	F(m) := \int_{-\infty}^{\infty} \diff x \, x p_s(x,m), \label{eq:fm}
\end{align}
satisfying the symmetry
\begin{align}
	F(m) = & - F(-m).
	\label{eq:symmetry}
\end{align}
The self-consistency equation \eqref{eq:scc} now reads
\begin{align}
m = F(m) \label{eq:selfcons2}
\end{align}
which obviously has a solution $m=0$. For $m>0$ the curvature of $F(m)$ is negative as shown in \cite{Da83} using a simple version of the Griffiths-Hurst-Sherman inequality \cite{EN76}. If the derivative of $F$ at $m=0$ is larger than one, there exists exactly one positive solution to Eq.~\eqref{eq:selfcons2} \cite{Da83}. Then, by symmetry we also have a negative solution. Otherwise $m=0$ is the only solution. Shiino \cite{Sh85,Sh87} showed that the solutions with $m \neq 0$ are stable if they exist whereas the $m=0$ solution is unstable in that case. The phase transition condition
\begin{align}
\partial_m F(m, a=a_c)|_{m=0} = 1.
\label{eq:ptc1}
\end{align}
can be written as 
\begin{align}
	\phi (a, D, \sigma) = 0 ,
	\label{eq:ptcphi}
\end{align}
where
\begin{align}
	\phi(a, D, \sigma) := \frac{  \frac{2D}{\sigma^2}\int_{-\infty}^{\infty}\diff x \, x^2 \exp \left[ \frac{2}{\sigma^2} (\frac{a-D}{2}x^2 - \frac{1}{4} x^4) \right] }{ \int_{-\infty}^{\infty}\diff x \, \exp \left[ \frac{2}{\sigma^2} (\frac{a-D}{2}x^2 - \frac{1}{4} x^4) \right]} -1.
	\label{eq:varphi}
\end{align}
This defines the critical manifold in the space spanned by $(a, D, \sigma)$. We observe immediately that on the critical manifold we have
\begin{align}
	\langle x^2 \rangle |_{\text{crit}} = \frac{\sigma^{2}}{2D}.
	\label{eq:secondmomentcrit}
\end{align}

In the following we show that the critical manifold is well-behaved: given any two of the parameters $a, D$ or $\sigma$, there exists a unique value, the critical value, of the third parameter which solves Eq. \eqref{eq:ptcphi}. For $D\le 0$ there is no solution to \eqref{eq:ptcphi}. Therefore we consider $D>0$ and furthermore $\sigma>0$ since there are only contributions in $\sigma^2$, negative $\sigma$ is equivalent. $\phi$ is continuous and continuously differentiable in $a, D$ and $\sigma$. $\phi$ is even $C^{\infty}$ on $\mathbb{R}\times \mathbb{R}_{+}\times \mathbb{R}_{+}$. By asymptotic evaluation of the integrals in \eqref{eq:varphi} we find
\begin{align}
	\lim_{a \rightarrow \infty} \phi(a, D, \sigma) &= + \infty , \\
	\lim_{a \rightarrow -\infty} \phi(a, D, \sigma) &= -1.
	\label{eq:limitsphi}
\end{align}
Because of continuity, for every $D, \sigma > 0$ there exists an $a$ satisfying Eq. \eqref{eq:ptcphi}. Since
\begin{align}
\partial_a \phi(a,D,\sigma) = \frac{2D}{\sigma^4} \left( \langle x^4 \rangle - \langle x^2\rangle^2 \right) >0,
\label{eq:monotoneina}
\end{align}
this solution is unique, $a=a_c$. The critical parameter $a_c$ is bounded by
\begin{align}
	\frac{1}{2}\frac{\sigma^2}{D} < a_c < \frac{3}{2} \frac{ \sigma^2}{D},
	\label{eq:ineqac}
\end{align}
as proven in Appendix \ref{sec:proofac}. These are the best possible bounds since the upper and the lower bound are asymptotically reached for weak or strong noise, respectively, as shown in Sects. 
\ref{subsec:wn} and \ref{subsec:sn} below.

Since $a_c>0$ we consider $\phi$ on $\mathbb{R}_{+}^{3}$. Because of Eq.~\eqref{eq:monotoneina} we can apply the implicit function theorem: there is locally around a solution of Eq.~\eqref{eq:ptcphi} a unique $C^{\infty}$ function $f_a(D, \sigma): \mathbb{R}_{+}\times \mathbb{R}_{+} \rightarrow \mathbb{R}_{+}$ satisfying
\begin{align}
	\phi(f_a(D, \sigma), D, \sigma) = 0.
	\label{eq:localfa}
\end{align}
Since there exists a unique solution $a_c = f_a(D, \sigma)$ to Eq.~\eqref{eq:ptcphi} for every $D, \sigma > 0$ the function $f_a(D, \sigma)$ is globally uniquely defined on $\mathbb{R}_{+}^{2}$ and is in $C^{\infty}$. Looking at relation \eqref{eq:ineqac} we see furthermore that $a_c= f_a(D, \sigma)$ assumes all values in $\mathbb{R}_{+}$ even if one of the variables $D$ or $\sigma$ is fixed to some value. We consider a fixed $\sigma = \sigma_0$ and exploit that $\phi(f_a(D, \sigma_0), D, \sigma_0) = 0 $ for all $D$. Therefore
\begin{align}
0 =&	\frac{\diff}{\diff D} \phi(f_a(D, \sigma_0), D, \sigma_0) \notag \\
=& \frac{2}{\sigma_0^2} \langle x^2 \rangle |_{a_c} - \frac{2D}{\sigma_0^4}( \langle x^4 \rangle |_{a_c} - \langle x^2 \rangle ^2 |_{a_c}  )(1- \frac{\partial f_a}{\partial D}) .
	\label{eq:calcddfa}
\end{align}
Solving for $\partial_{D}f_a$ and inserting the relations \eqref{eq:secondmomentcrit} and \eqref{eq:xfourac} we find
\begin{align}
	\frac{\partial}{\partial D} f_a(D, \sigma_0) = \frac{a_c - \frac{3 \sigma_{0}^2}{2D} }{a_c - \frac{\sigma_{0}^2}{2D}} < 0,
	\label{eq:monotonfaind}
\end{align}
where the negativity is guaranteed by inequality \eqref{eq:ineqac}. Because of the monotonicity and surjectivity of $f_a(D, \sigma_0)$ there exists an inverse function $f_D(a, \sigma_0): \mathbb{R}_{+} \rightarrow \mathbb{R}_{+}$ for all $\sigma_0 >0$, i.e., we have the function $f_D(a, \sigma): \mathbb{R}_{+}\times \mathbb{R}_{+} \rightarrow \mathbb{R}_{+}$ satisfying 
\begin{align}
	\phi(a, f_D(a, \sigma), \sigma) = 0.
	\label{eq:fdzero}
\end{align}
Analogously, for fixed $D = D_0$ we calculate
\begin{align}
	0 = &\, \frac{\diff}{\diff \left(1/ \sigma^2\right)} \phi(f_{a}(D_0, \sigma), D_0, \sigma) \notag \\
	= &\, 2D_0 \langle x^2 \rangle + \frac{2D_0}{\sigma^2} \left[(f_a-D_0) \langle x^4 \rangle - \frac{1}{2} \langle x^6\rangle \right] \notag \\
	& - \frac{2D_0}{\sigma^2} \left[ (f_a - D_0)\langle x^2\rangle^2 - \frac{1}{2}\langle x^2 \rangle \langle x^4 \rangle \right] \notag \\
	& + \frac{2D_0}{\sigma^4}\left( \langle x^4 \rangle - \langle x^2 \rangle^2 \right) \frac{\partial f_a}{\partial \left( {1}/{\sigma^2}\right)}.
	\label{eq:calcdsigmafa}
\end{align}
Inserting the expressions \eqref{eq:secondmomentcrit}, \eqref{eq:xfourac} and \eqref{eq:xsixac} we find
\begin{align}
	\frac{\partial f_a}{\partial \left( {1}/{\sigma^2}\right)} = -\frac{\sigma^2}{2} \label{eq:monotonfainsigma1} 
	 \left( D_0 \frac{\frac{3\sigma^2}{2D_0}- a_c  }{a_c - \frac{\sigma^2}{2D_0}} + a_c \right) < 0, 
\end{align}
where inequality \eqref{eq:ineqac} was used. Hence we find
\begin{align}
	\frac{\partial}{\partial \sigma} f_a(D, \sigma) > 0.
	\label{eq:monotonfainsigma2}
\end{align}
From the monotonicity there follows the existence of a function $f_{\sigma}(a,D)$ satisfying
\begin{align}
	\phi(a, D, f_{\sigma}(a,D)) = 0
	\label{eq:fsigmazero}
\end{align}
in analogy to the previous case.\\ 
Hence, given any two parameters, positive, there exists a unique critical value of the third parameter denoted by $f_a$, $f_D$, or $f_{\sigma}$. Because of the monotonicity of $\phi$ with respect to $a$ and of $f_a$ with respect to $D$ and $\sigma$ we conclude that there exists a pair of nonzero solutions $m_{\pm}$ of Eq.~\eqref{eq:selfcons2} if and only if one of the  following equivalent conditions is satisfied 
\begin{align}
	a >& f_{a}(D, \sigma), \\
	D >& f_{D}(a, \sigma), \\
	\sigma <& f_{\sigma}(a, D).
	\label{eq:bifurcationpoints}
\end{align}

\subsection{Scaling\label{subsec:scaling}}
\indent With an arbitrary $\tau > 0$ we can rescale variables and parameters
as 
\begin{align}
t' &=  \tau t , \hspace{2cm}	&x_i' &=\tau^{-1/2} x_i,
\label{eq:trafo2}\\
a'& =  \tau^{-1}a,    &D' &= \tau^{-1} D, \label{eq:trafo4}\\
\xi_{i} ' (t')& =  \tau^{-3/2} \xi_{i}(t),   &\sigma' &=
\tau^{-1} \sigma, \label{eq:trafo6}
\end{align}
which leads to a system of Langevin equations equivalent to Eqs.~\eqref{eq:eqofmotion}
\begin{align}
	\frac{\diff }{\diff t'} x_i ' =  a'x_i' - x_i'^3 - \frac{D'}{L-1} \sum_{j=1}^{L} (x_i'-x_j') + \xi_i'(t'),
\end{align}
for $i=1,\dots,L$, with
\begin{align}
\langle\xi_{i}'\left(t'\right)&\xi_{j}'\left(s'\right)\rangle = \sigma'^{2}\delta_{ij}\delta\left(t'-s'\right).
	\label{eq:langtrafo1}
\end{align}
Hence we have not three but only two independent parameters. In a similar way, also the general case with a coefficient of the cubic term in Eqs.~\eqref{eq:eqofmotion} can be treated, ending with only two independent parameters. 

\begin{figure}[H]
\includegraphics{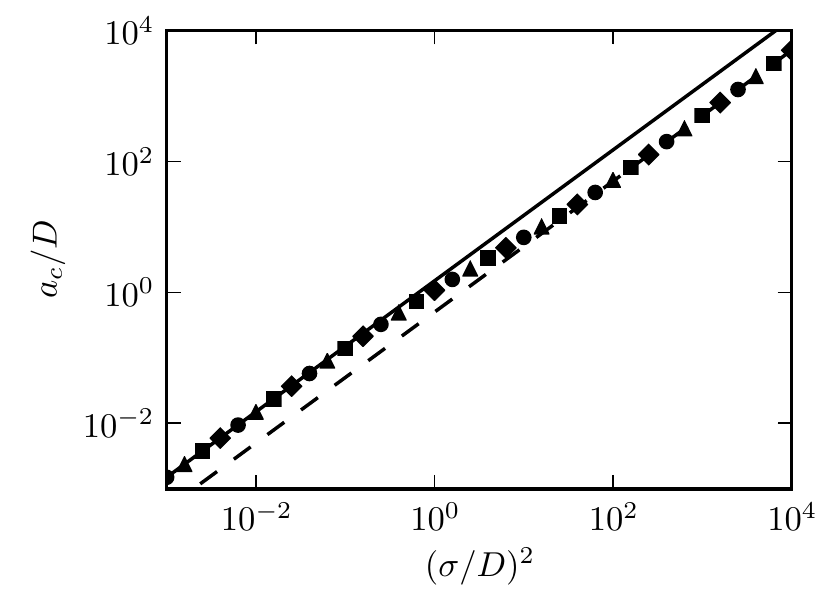}
\caption{Critical parameter $a_c/D$ in dependence of noise strength $(\sigma/D)^2$ as a log-log plot. Here the equation \eqref{eq:ptcphi}
was numerical solved
for $D=10^{-3}$ $\left(\blacktriangle\right)$, $D=10^{-2}$ $\left(\bullet\right)$,
$D=10^{-1}$ $\left(\blacklozenge
\right)$, and $D=1$ $\left(\blacksquare\right)$, each for several values of $\sigma$.
The solid line describes the function $3\sigma^{2}/(2D^{2})$, the limit of weak noise, the dashed line the function
$\sigma^{2}/(2D^{2})$, the asymptote for strong noise.\label{fig:Log-Log-Plot}
}
\end{figure}

We observe that $\phi(a, D, \sigma)$ given by Eq. \eqref{eq:varphi} is invariant under this rescaling. This allows the following argument. We set $\tau=D$ such that $D'=1$ and find $a_{c}'=f_{a}(1, \sigma')$. Therefore
\begin{align}
	a_c = f_a(D, \sigma) = D f_a\left(1, \frac{\sigma}{D}\right).
	\label{eq:acscaled}
\end{align}
Fig. \ref{fig:Log-Log-Plot} shows $a_c/D$ as a function of $\left(\sigma/D\right)^2$ in a log-log plot for different values of $D$ as obtained by numerical solution of Eq. \eqref{eq:ptcphi}. It confirms that $a_c/D$ depends only on the ratio of $\sigma$ and $D$ as predicted by Eq.~\eqref{eq:acscaled}. Therefore $a_c(\sigma, D)$ exhibits the same asymptotic behavior for strong coupling as for weak noise, and for weak coupling as for strong noise, respectively. 

Dawson \cite{Da83} uses a scaling where $a'=1$ and claims that, given $D'$, the critical noise strength $\sigma_c'$ is bounded by $2^{-1/2}< \sigma_c'/D'< 2^{1/2}$, cf. Eq. $(3.43)$ there. We show that this assertion is not true. From Eqs.  \eqref{eq:trafo4} and \eqref{eq:trafo6} we have ${\sigma_c'}/{D'}= {\sigma_c}/{D} = f_{\sigma}(a, D)/D$, where $f_{\sigma}(a,D)$ is defined by Eq. \eqref{eq:fsigmazero}. We now choose $D=1$ and $a= f_a(\sigma=\Sigma, D=1)$, where $\Sigma$ can have any positive value and $f_a$ is defined by Eq. \eqref{eq:localfa}. Since $f_{\sigma}(f_a(\Sigma, 1), 1) = \Sigma$ it holds that $\sigma_{c}'/D'= \Sigma$ which can be chosen beyond the bonds asserted in \cite{Da83}. However the property that for any $D'>0$ there exists a unique critical $\sigma_c'$ remains true, as we have shown in Sec. \ref{subsec:cmgeneral}.

\begin{figure}[h]
	\centering
	{	\includegraphics{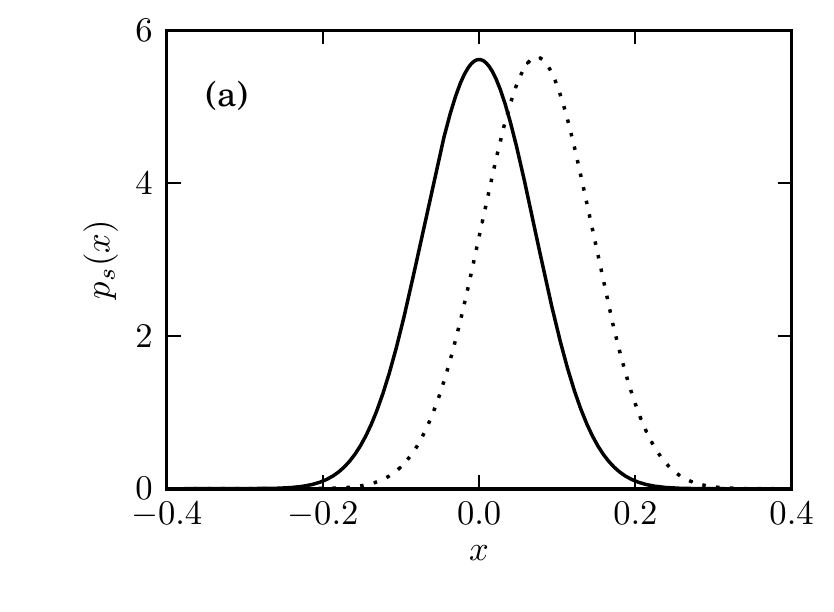}
	\includegraphics{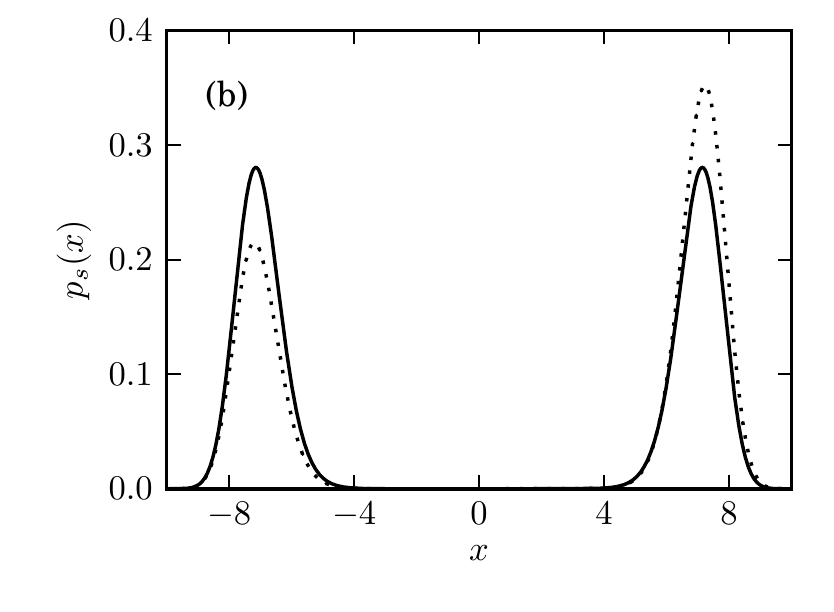}}
	\caption{Mechanism of symmetry breaking for weak and for strong noise. Stationary distribution $p_s(x)$ at the critical point (solid line) and just above (dotted line). (a) For weak noise $(\sigma=0.1)$ the distribution is Gaussian, centered around zero $(m=0)$ for $a=a_c=0.0149$, and at $a=0.02$ rigidly shifted to the right $(m=0.073)$. (b) For strong noise $(\sigma=10)$ the distribution is bimodal with sharp peaks, centered around zero, with equal weights for $a=a_c=52.04$ and with different weights for $a=53$ $(m=1.71)$. Coupling strength $D=1$.\label{fig:symbreak}}
\end{figure}

At $a= a_c(D, \sigma)$ the stationary probability density $p_s(x)$ for the
coordinate of an arbitrary  constituent is qualitatively different for weak and
for strong noise, cf. Fig. \ref{fig:symbreak}.
For weak noise $p_s(x)$ is approximately a Gaussian centered
at $x=0$, whereas for strong noise it is the sum of two equally weighted narrow
peaks located at $\pm \sigma \sqrt{1/D}$. For $a>a_c$ the symmetry is broken in
different ways. In the weak noise limit $p_s(x)$ is still a Gaussian but
centered at $\langle x \rangle \ne 0$, whereas in the strong noise limit the two
narrow peaks stay at $\pm \sigma \sqrt{1/D}$ but their weights become unequal
such that $\langle x \rangle \ne 0$ also in this case. 
The critical parameter $a_c(D, \sigma)$ is bounded between its limit values for
strong and weak noise respectively, as will be shown rigorously in Appendix \ref{sec:proofac}.

Exploiting that at criticality the even moments are explicitly known we give a simple
handwaving argument which leads to the correct leading behavior for weak and strong noise. Especially we use Eq.~\eqref{eq:ptcphi}, that is
$\langle x^2\rangle_{\text{crit}}= \sigma^2/(2D)$ and, cf. Appendix \ref{sec:moments}, $\langle
x^4\rangle_{\text{crit}}= a_c \langle x^2\rangle_{\text{crit}}$. The fourth
cumulant, the kurtosis, of a Gaussian is zero and therefore $\langle x^4\rangle
= 3 \langle x^2\rangle^2$. Comparing this with the above expressions we obtain
for weak noise in leading order $a_c = 3\sigma^2 / (2D)$, which is the upper bound in the inequality \eqref{eq:ineqac}. Furthermore, for a
symmetric probability density of two narrow peaks, the variance of $x^2$ is
approximately zero and therefore $\langle x^4\rangle = \langle  x^2\rangle^2$.
Comparing with the above expressions we obtain for strong noise in leading order
$a_c = \sigma^2/(2D)$, which is the lower bound for $a_c$ in \eqref{eq:ineqac}. 

In Eqs. \eqref{eq:monotonfaind} and \eqref{eq:monotonfainsigma1} we have ordinary differential equations for $a_c$ as a function of $D$ and $1/\sigma^2$. By substituting $\alpha = a_c/D$ and $\beta = \sigma^2/D^2$ we find from either of these equations
\begin{align}
	\frac{\diff \alpha}{\diff \beta} = \frac{\alpha}{2\beta} + \frac{1}{2\beta} \, \frac{3/2- \alpha/\beta}{\alpha/\beta - 1/2}.
	\label{eq:ode}
\end{align}
In the next two sections we will systematically study the behavior of the critical parameter for weak and strong noise using Eq.~\eqref{eq:ode}. The same results can be obtained by asymptotic evaluation of the integrals in the phase transition condition \eqref{eq:ptcphi} as shown in Appendix \ref{sec:laplace}.

\subsection{Weak Noise\label{subsec:wn}}
\indent Because of inequality~\eqref{eq:ineqac} it holds that
\begin{align}
	a_c \rightarrow 0 \quad \text{for} \quad \sigma^2 \rightarrow 0.
	\label{eq:a_cconvwn}
\end{align}
That means we can continuously extend $a_{c}$ to $\sigma = 0$. To obtain the asymptotic behavior of $a_{c}$ for weak noise we make the ansatz
\begin{align}
	a_c(\sigma, D)/D = \alpha = \sum_{n=1}^{\infty} a_{n} \beta^{n}  \label{eq:ansatzwn}.
\end{align}
Inserting this series in the differential equation \eqref{eq:ode} and comparing coefficients in powers of $\beta$ we find the recursion relation
\begin{align}
	a_{n+1} =& \left(n- \frac{1}{2} \right) a_{n} - \sum_{k=1}^{n} \left[2(n-k) +1 \right]a_{n-k+1}a_{k} 
	\label{eq:coeffweaknoise}
\end{align}
for $n=1,2, \dots$ with initial condition $a_1=3/2$.
For $\alpha$ the three leading terms as $\beta \rightarrow 0$ are
\begin{align}
\alpha =  \frac{3}{2} \beta - \frac{3}{2}\beta^2 + \frac{27}{4} \beta^3  + \mathcal{O}(\beta^4) \label{eq:weaknoiseac2} 
\end{align}
which coincides with the result in Appendix \ref{sec:laplacewn}. We observe that the upper bound in \eqref{eq:ineqac} is reached asymptotically.

\begin{figure}[H]
	\includegraphics{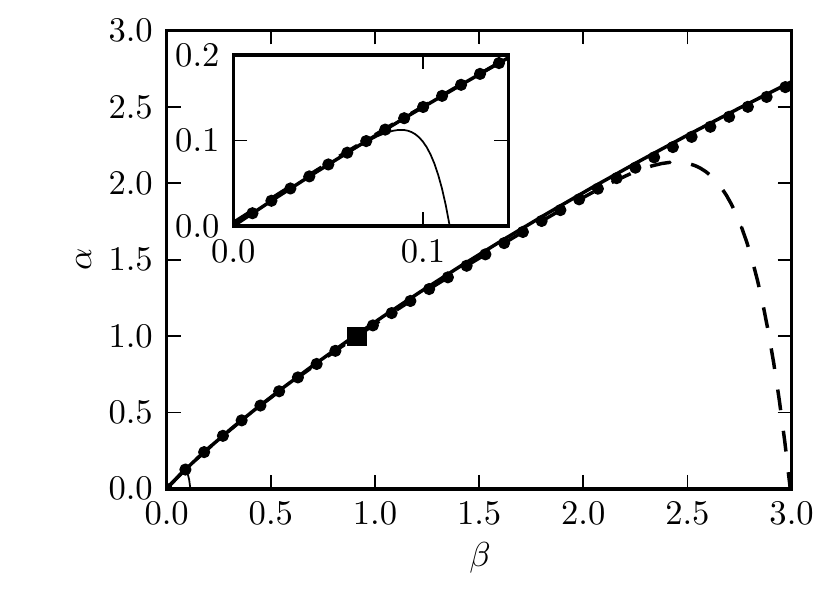}
	\caption{Critical parameter $\alpha$ as a function of $\beta$. The point with $\alpha=1$ is known exactly $(\blacksquare)$. The figure shows the numerical solution $(\bullet)$ of Eq. \eqref{eq:ptcphi}, the series for weak noise up to $\beta^{10}$ (thin solid line, see also insert), the corresponding Pad\'e approximant $p_{10, 10}$ (thick solid line), and the series around the exactly known point $(\blacksquare)$ up to 10th order (dashed line).\label{fig:weaknoiselin}}
\end{figure}

\begin{figure}[H]
	\includegraphics{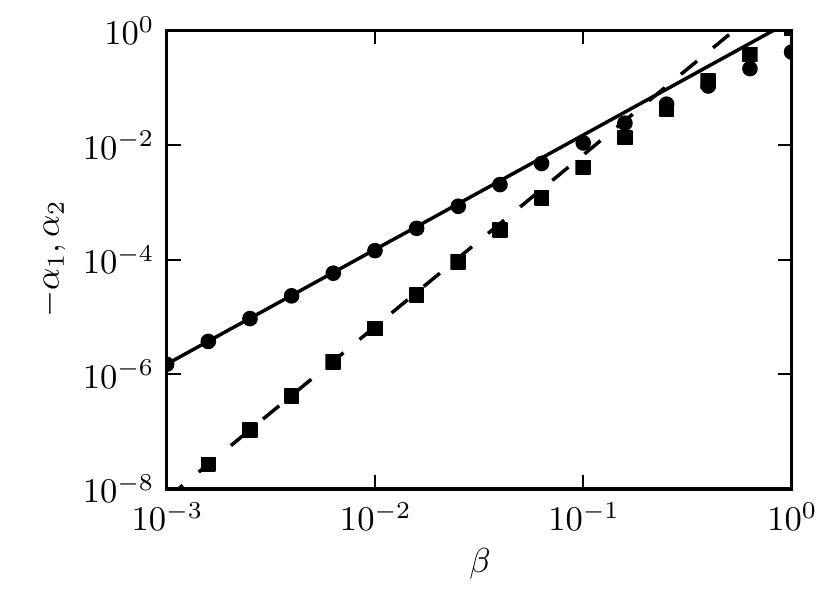}
	\caption{Test of the scaling of $a_c$ for small $\beta$. The symbols show the results for $-\alpha_1(\bullet)$ and $\alpha_2(\blacksquare)$ obtained from the numerical solution of Eq. \eqref{eq:ptcphi}. The predicted behavior $-\alpha_1 \sim 3/2 \beta^2$ (solid line) and $\alpha_2 \sim 27/4 \beta^3$ (dashed line) is confirmed.\label{fig:numasympwn}}
\end{figure}

In Fig.~\ref{fig:weaknoiselin} we see $\alpha$ as a function of $\beta$. The series \eqref{eq:ansatzwn} up to $\beta^{10}$ coincides with the numerical solution of Eq. \eqref{eq:ptcphi} only for very small values of $\beta$ and taking into account more terms does not seem to improve the result for larger values of $\beta$. The figure shows also the Pad\'e approximant $p_{10,10}$ which coincides much better with the numerical results. The Pad\'e approximant $p_{N, N}$ is a rational function $q_1/q_2$, where $q_1$ and $q_2$ are polynomials of degree $N$ and the Taylor series of $p_{N, N}$ agrees with the series \eqref{eq:ansatzwn} up to $\beta^N$ \cite{BO78}.
We have not been able to prove convergence of the series \eqref{eq:ansatzwn} near $\beta=0$. It is clear by its definition via the implicit function that $a_{c}(\sigma, D)$ is an analytic function for any positive $\sigma$, but at $\sigma=0$ we don't know. Nevertheless, Eq.~\eqref{eq:weaknoiseac2} has a meaning as it correctly describes the asymptotic behavior of $\alpha$ obtained by numerically solving the PTC \eqref{eq:ptcphi} for $\beta \rightarrow 0$ as $\alpha \sim 3/2\, \beta$ where the symbol $\sim$ means $\lim_{\beta \rightarrow 0}\alpha/\beta = 3/2$, cf. Fig. \ref{fig:Log-Log-Plot}. The coefficients of the higher order terms in Eq.~\eqref{eq:weaknoiseac2} give systematic corrections in the sense that $\alpha_1 := \alpha - 3/2\, \beta \sim -3/2 \, \beta^2$ and $\alpha_2 := \alpha_1 + 3/2\, \beta^2\sim 27/4 \, \beta^3$, cf. Fig. \ref{fig:numasympwn}.

\subsection{Strong Noise\label{subsec:sn}}
\begin{figure}[H]
	\includegraphics{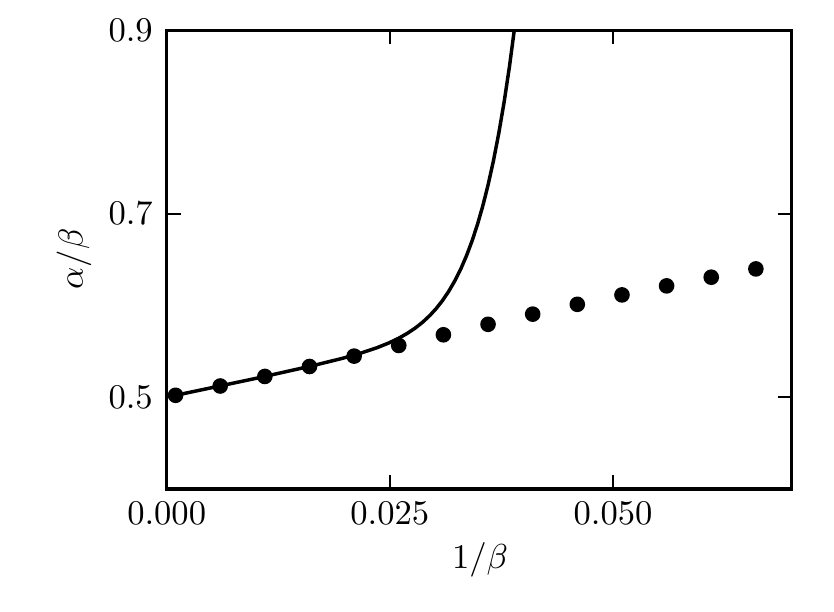}
	\caption{Critical parameter $\alpha/\beta$ as a function of $1/\beta$. The figure shows the numerical solution $(\bullet)$ of Eq.~\eqref{eq:ptcphi} compared with the series for strong noise (solid line) up to $(1/\beta)^{10}$.\label{fig:strongnoiseinv}}
\end{figure}

\begin{figure}[H]
	\includegraphics{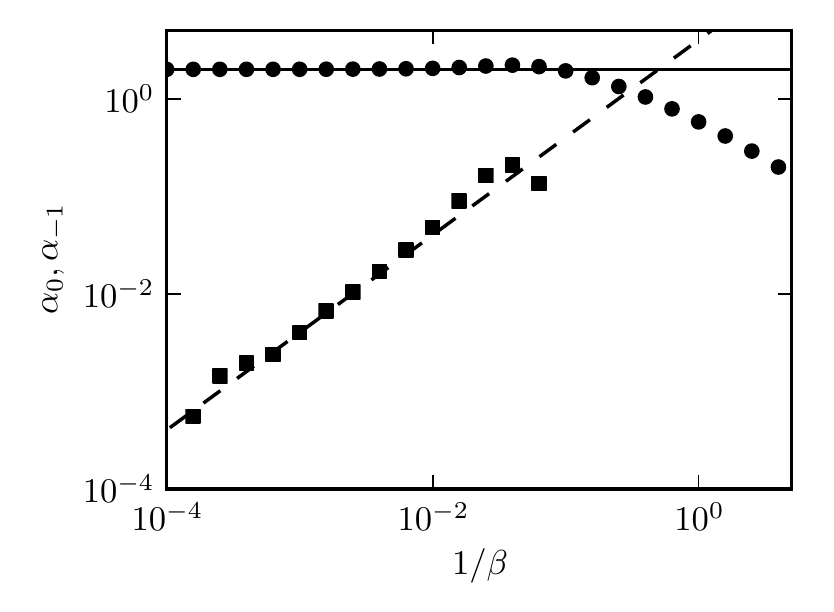}
	\caption{Test of the scaling of $a_c$ for large $\beta$. The symbols show the results for $\alpha_{0}(\bullet)$ and $\alpha_{-1}(\blacksquare)$ obtained from the numerical solution of Eq.~\eqref{eq:ptcphi}. The predicted behavior $\alpha_0 \sim 2 $ (solid line) and $\alpha_{-1}= 4/\beta$ (dashed line) is confirmed.\label{fig:numasympsn}}
\end{figure}

\indent In the limit of strong noise $\sigma \rightarrow \infty$, again motivated by \eqref{eq:ineqac}, we use a series ansatz for $\alpha$ as $\sigma^2 \rightarrow \infty$,
\begin{align}
	a_c/D = \alpha = a_{1}\beta + \sum_{i=0}^{\infty} a_{-i} \beta^{-i} \label{eq:ansatzsn}.
\end{align}
Inserting in Eq.~\eqref{eq:ode} and comparing coefficients in powers of $\beta$ leads to
\begin{align}
	a_{1} = \frac{1}{2}, \qquad a_{0} = 2,
\end{align}
and to the backward recursion
\begin{align}
	a_{-(n+1)}=& -2a_{-n} \notag \\
	& +4\sum_{k=0}^{n}(n-k+\frac{1}{2})a_{-k} a_{-(n-k)} 
	\label{eq:coeffstrongnoise}
\end{align}
for $n=0,1,2, \dots$ with initial condition $a_0=2$. The three leading terms for $\alpha$ as $\beta \rightarrow \infty$ are
\begin{align}
	\alpha = \frac{1}{2} \beta + 2 + 4\frac{1}{\beta} + \mathcal{O}\left( \frac{1}{\beta^2}\right), \label{eq:strongnoiseac}
\end{align}
The same coefficients are obtained via Laplace's method, cf. Appendix~\ref{sec:laplacesn}. In the case of strong noise we reach the lower bound in \eqref{eq:ineqac} asymptotically.
In Fig. \ref{fig:strongnoiseinv} we see $\alpha/\beta$ as a function of $1/\beta$, where the numerical solution of Eq.~\eqref{eq:ptcphi} is compared with the series \eqref{eq:ansatzsn}. The series agrees with the numerical results only for very small values of $1/\beta$. Here, also the Pad\'e approximants do not work as well as for weak noise, since they have many poles within the region of interest.
As for weak noise we have not been able to prove convergence of the series \eqref{eq:ansatzsn}. Nevertheless, Eq.~\eqref{eq:strongnoiseac} describes the asymptotics of $\alpha$ for $\beta \rightarrow \infty$ as $\alpha \sim 1/2 \, \beta$, cf. Fig. \ref{fig:Log-Log-Plot}, and the higher order coefficients in Eq.~\eqref{eq:strongnoiseac} give systematic corrections in the sense $\alpha_0:=\alpha-1/2 \, \beta \sim 2$, and $\alpha_{-1} := \alpha_0 - 2 \sim 4/\beta$, cf. Fig. \ref{fig:numasympsn}.

\subsection{An Intermediate Regime}

Dawson \cite{Da83} observed that, in our notation, considering only the sub-manifold of the critical manifold defined by the condition $a=D$, it is possible to obtain an explicit expression for the critical value of $a$ as a function of $\sigma$. Substituting $y=x/\sqrt{\sqrt{2}\sigma}$ in Eq. \eqref{eq:varphi} and solving Eq. \eqref{eq:ptcphi} for $a$ under the restriction that $a=D$ yields the critical parameter
\begin{align}
	a_{c}(D=a, \sigma) & =  2^{-3/2} \frac{\int_{-\infty}^{\infty} \diff y \, \exp \left(-y^4 \right)}{ \int_{-\infty}^{\infty} \diff y \, y^{2}\exp \left(-y^4 \right)} \sigma \notag \\
	& =  2^{-3/2} \frac{\Gamma(1/4)}{\Gamma(3/4)} \sigma \approx 1.046 \sigma.
	\label{eq:submanifold}
\end{align}
In terms of the rescaled parameters $\alpha = a_{c}/D$ and $\beta = \sigma^{2}/D^{2}$ this reads
\begin{align}
	\alpha\left(\beta_{0}= \frac{8\Gamma(3/4)^{2}}{\Gamma(1/4)^{2}}\right) = 1.
	\label{eq:submanifoldrescaled}
\end{align}
Using the ansatz
\begin{align}
	\alpha(\beta) = 1 + \sum_{i=1}^{\infty} a_{i} (\beta - \beta_{0})^{i}
	\label{eq:seriesintermediate}
\end{align}
we find with the ODE \eqref{eq:ode} the recursion formula
	  \begin{align}
		  a_{n+1}=& \frac{1}{(n+1)(\beta_{0}^{2}/2-\beta_{0})}\Big\{ a_{n} \big(n+ \beta_0 a_1 -1/2 - \beta_0 n \notag\\
		  + & 1/4\beta_0\big)  + a_{n-1}(-1/2n +3/4) \notag\\
		  &+ \sum_{k=1}^{n-1}\big[(n-k-1/2)a_k a_{n-k} \notag\\
			  &+ \beta_0 a_k a_{n-k+1}(n-k+1)\big]\Big\},
	  \end{align}
for $n=2,3, \dots$ with initial conditions 
	  \begin{align}
		  a_{1} = \frac{1}{2-\beta_{0}}, \quad a_{2} = \frac{1-\beta_{0}-\beta_{0}^{2}/4}{\beta_{0}(2-\beta_{0})^{3}}.
	  \end{align}
	  In Fig. \ref{fig:weaknoiselin} we see good agreement between the numerical solution of the phase transition condition \eqref{eq:ptcphi} and the series \eqref{eq:seriesintermediate} up to the tenth order term.

\section{Critical Behavior \label{sec:critbeh}}

\subsection{Order Parameter \label{subsec:orderparam}}

\indent To calculate the behavior of the order parameter $m$ for $a$ close to the critical value $a_c$ it is convenient to introduce the notation
\begin{align}
	N_k(m, a, D, \sigma) := & \int_{-\infty}^{\infty} \diff x \, x^k \label{eq:nk} \\
& \times \exp\left[\frac{2}{\sigma^2} (mDx + \frac{a-D}{2}x^2 -\frac{1}{4}x^4) \right]. \nonumber
\end{align}
$F(m)$ defined in Eq.~\eqref{eq:fm} can be expressed as
\begin{align}
	F(m) = \frac{N_{1}}{N_{0}} = \frac{\sigma^2}{2D} \partial_{m} \ln N_{0}(m, a, D, \sigma)
	\label{eq:fm2}
\end{align}
and the $k$th moment of the probability density $p_s(x,m)$ as
\begin{align}
\langle x^k \rangle = \frac{N_{k}(m, a, D, \sigma)}{N_{0}(m, a, D,\sigma)}.
\end{align}
Expanding the right hand side of Eq.~\eqref{eq:fm2} for small $m$, according to \eqref{eq:selfcons2} we obtain the self-consistency equation
\begin{align}
m = & \left( \frac{2D}{\sigma^2} \right) \langle x^2 \rangle_{0} \, \, m \notag\\
& + \left(\frac{2D}{\sigma^2}\right)^3 \cdot \left( \frac{\langle x^4 \rangle_{0}}{6}-\frac{\langle x^2 \rangle^2_{0}}{2} \right) m^3 + \mathcal{O}(m^{5}), \label{eq:sccsmallm}
\end{align}
where $\langle x ^k \rangle_{0} = \langle x^k \rangle |_{m=0}$.

\noindent Equation \eqref{eq:sccsmallm} has always the trivial solution $m=0$. For $a > a_c$ there is a pair of nontrivial real solutions
\begin{align}
	m_{\pm} = \pm  \frac{\sigma^2}{2D}\sqrt{\frac{\langle x^2 \rangle_{0}-\frac{\sigma^2}{2D}}{\langle x^2 \rangle^2_{0}/2 - \langle x^4 \rangle_{0}/6 }} \label{eq:critm}
\end{align}
since the denominator of the radicand is always positive as proven in Appendix \ref{sec:proofac} and the numerator of the radi\-cand is positive if and only if $a> a_c$. This follows from the monotonicity of the second moment as a function of $a$, cf. inequality \eqref{eq:monotoneina}. Hence the expansion \eqref{eq:sccsmallm} is sufficient to determine the leading behavior of $m$ close to the critical point and we can exclude the existence of a tricritical point.

\begin{figure}[H]
\includegraphics{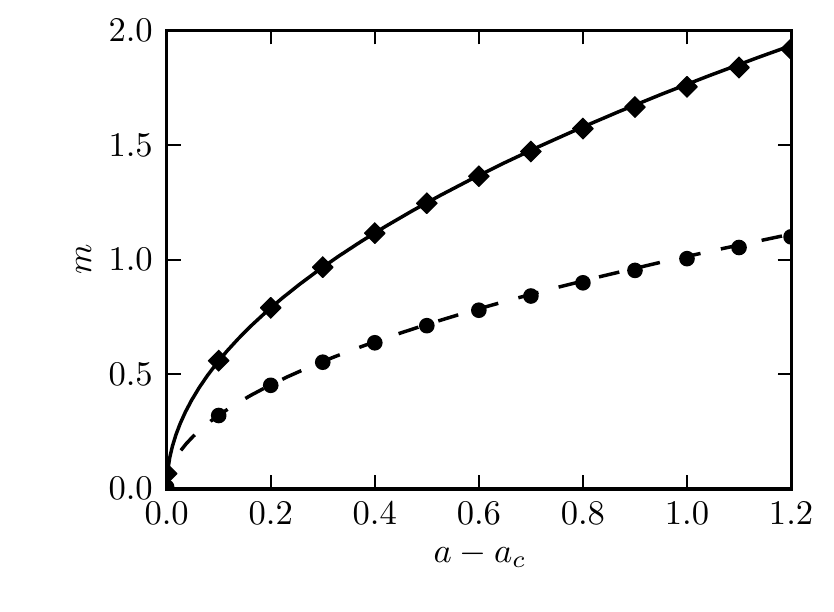}
\caption{Critical behavior of the order parameter $m$ for \mbox{$\sigma=10\,(\blacklozenge)$} and $\sigma=0.1\,(\bullet)$ determined by numerical evaluation of Eq.~\eqref{eq:selfcons2} compared to the analytical results from Eqs. \eqref{eq:critweaknoise} and \eqref{eq:critstrongnoise} for weak noise (dashed line) and for strong noise (solid line). Coupling strength $D=1$.\label{fig:2}}
\end{figure}

We now expand the right hand side of Eq. \eqref{eq:critm} for small $a-a_c=\varepsilon>0$, exploiting that at $a=a_c$ all even moments can be determined recursively from Eq.~\eqref{eq:secondmomentcrit}, see Appendix \ref{sec:moments}. Inserting Eqs. \eqref{eq:secondmomentcrit}, \eqref{eq:xfourac} and \eqref{eq:xsqcrit} in Eq. \eqref{eq:critm} yields in leading order
\begin{align}
	m_{\pm} = \pm \frac{\sqrt{6}\, \sigma}{2D}\sqrt{\frac{a_c- \frac{\sigma^{2}}{2D}}{\frac{3\sigma^{2}}{2D}-a_c}} \varepsilon^{1/2}.
	\label{eq:critbehav}
\end{align}
Hence we have found the typical mean field exponent $1/2$ and an analytic expression for the amplitude in terms of the critical parameter. Inserting $a_c$ in the limit of weak noise from Eq.~\eqref{eq:weaknoiseac2} leads to
\begin{align}
	m_{\pm} = & \pm \left(1 + \frac{3}{2} \frac{\sigma^2}{D^2} + \mathcal{O}(\sigma^4) \right) \varepsilon^{1/2}. \label{eq:critweaknoise}
\end{align}
In the limit of strong noise we obtain with Eq.~\eqref{eq:strongnoiseac} 
\begin{align}
	m_{\pm} = & \pm \sqrt{3} \left( 1 +2 \frac{D^2}{\sigma^2}+\mathcal{O}\left(\frac{1}{\sigma^4}\right) \right) \varepsilon^{1/2}. \label{eq:critstrongnoise}
\end{align}
In Fig.~\ref{fig:2} these analytical results are compared for a typi\-cal parameter setting with the numerical evaluation of the self-consistency equation \eqref{eq:selfcons2}.

\subsection{Susceptibility\label{subsec:susc}}
In this section we observe that the susceptibility is diverging at the critical point as $\chi \sim A_{\pm}/(a-a_c)$ with the amplitudes $A_{+}$ and $A_{-}$ for $a>a_c$ and $a<a_c$, respectively. We explicitly calculate the amplitudes $A_{\pm}$ in terms of $a_c$ and find a universal ratio between them. The whole procedure, as well as the results, are in complete analogy to equilibrium thermodynamics. However the calculation is explicitly possible here, and up to our knowledge it has not been done in this context before.

\indent We introduce an external field $h$ in Eq.~\eqref{eq:eqofmotion}
\begin{align}
\dot{x_i} =  h + a x_i - x_i^3 - \frac{D}{L-1}\sum_{j=1}^{L}(x_i -x_j) + \xi_i(t).
\end{align}
The susceptibility is defined as the response of the system to a small external field
\begin{align}
\chi =  \frac{\partial m}{\partial h} \bigg|_{h=0}.
\end{align}
We investigate the susceptibility close to the critical point. In analogy to Eq.~\eqref{eq:sccsmallm} we find the self-consistency equation for small $m$ and $h$
\begin{align}
m = & \left( \frac{2D}{\sigma^2} \right) \langle x^2 \rangle_{0} (m+h)  \nonumber \\
& +\left(\frac{2D}{\sigma^2}\right)^3 \left( \frac{\langle x^4 \rangle_{0}}{6}-\frac{\langle x^2 \rangle^2_{0}}{2} \right)(m+h)^3 . \label{eq:sccsmallmh}
\end{align}
Taking the derivative with respect to $h$ at $h=0$ we obtain
\begin{align}
\chi  = & (\chi + 1)\frac{2D}{\sigma^2} \notag  \\
& \times \left( \langle x^2 \rangle_0 + m^2\frac{2D^2}{\sigma^4} \big(\langle x^4 \rangle_0 - 3 \langle x^2 \rangle_{0}^2 \big) \right). \label{eq:chisusc}
\end{align}
At the critical point we find
\begin{align}
\chi =  \chi + 1,
\end{align}
which can be satisfied only asymptotically by $\chi \rightarrow \pm \infty$.

Below the critical point we have $m=0$ and \mbox{$\langle x^2 \rangle_0 <\sigma^2/(2D)$}. Therefore \eqref{eq:chisusc} becomes
\begin{align}
	\chi = & \frac{\langle x^2 \rangle_{0}}{\frac{\sigma^{2}}{2D}- \langle x^2 \rangle _{0}}.
\end{align}
For small $ a - a_c = \varepsilon < 0$ we find the leading behavior of the susceptibility close to the critical point using Eq.~\eqref{eq:xsqcrit}
\begin{align}
	\chi =  A_{-}\frac{1}{-\varepsilon}\quad \text{with} \quad A_{-} = \frac{\sigma^{2}}{a_c - \frac{\sigma^{2}}{2D}}.
\end{align}

For $a - a_c = \varepsilon > 0 $ we obtain in leading order with the help of Eqs.~\eqref{eq:critbehav}, \eqref{eq:chisusc} and \eqref{eq:xsqcrit}
\begin{align}
	\chi =  A_{+} \frac{1}{\varepsilon} \quad \text{with} \quad A_{+} = \frac{1}{2} \frac{\sigma^{2}}{a_c - \frac{\sigma^{2}}{2D}}.
\end{align}
Since $A_{+} = A_{-}/2$ we have
\begin{align}
\lim_{\varepsilon\rightarrow +0} \frac{\chi(a_c-\varepsilon)}{\chi(a_c +\varepsilon)} = 2,
\end{align}
which is universal, i.e. not depending on parameters.

\section{Conclusion \label{sec:conclusion}}

In this paper we have proved an upper and a lower bound for the
critical parameter $a_c$. These bounds are optimal since they are
asymptotically reached for weak and for strong noise, respectively. We found an ordinary differential equation describing the critical point $a_c/D$ as a function of $(\sigma/D)^2$, which allows to explicitly give a recursion formula for all coefficients of the asymptotic expansion of $a_c$ for weak and strong noise as well as for an expansion around a special point, where $a_c$ is known exactly.

In the limits of weak and strong noise the mechanism of symmetry breaking is qualitatively different. For weak noise and $a$ close to $a_c$ the stationary distribution of the coordinates $p_s(x)$ is a
Gaussian. Below the critical point the Gaussian is centered around
zero. Above, for $a>a_c$, the Gaussian is shifted in positive or
negative direction, the symmetry is broken. For strong noise and $a$ close to $a_c$ $p_s(x)$ consists of two narrow
peaks located symmetrically with respect to zero. The symmetry is broken such that for $a>a_c$ one of the peaks
gains a larger weight than the other.

We have proved that the critical manifold is well behaved, that is if
two of the three positive parameters $a, D$, and $\sigma$ are given
there exists a unique critical value of the third. The proof hinges
on the knowledge of the above mentioned boundaries of $a_c$. The wellbehavedness of the critical manifold allows
to reduce the number of parameters and implies
certain scaling properties. For example, the ratio $a_c/D$ 
depends on noise strength and coupling constant only as a function of
$\sigma/{D}$,  and the limits of weak noise or strong coupling
and strong noise or weak coupling are equivalent.

We further have determined the critical behavior of order parameter
and susceptibility.  As well known, they follow as a function of
$a-a_c$ power laws with the mean field exponents. We have calculated
the amplitude of the order parameter in terms of the critical
parameter $a_c$ and explicitly in the limits of weak and strong noise
and found for the amplitude ratio of the susceptibilities the
universal law ${A_{-}}/ {A_{+}}=2$.

It is a natural question whether for systems with higher order nonlinearity similar results can be obtained. It is further desirable to study the critical manifold of a system with both additive and multiplicative noise.

\begin{acknowledgements}
\hyphenation{IMPRS}
Thanks is due to Klaus Kroy for valuable discussions. R. K. thanks the IMPRS Mathematics in the Sciences for funding.
\end{acknowledgements}

\appendix
\section{Moments of $x$\label{sec:moments}}
We need the exact recursive relations between moments which are derived by Dawson \cite{Da83} exploiting the Ito-formula. 
To keep the paper self-contained we rederive these relations in our notation using a different argument.

\indent By partial integration of the right hand side of Eq.~\eqref{eq:nk} we find
\begin{align}
N_k = & \int_{-\infty}^{\infty} \diff x \; x^k \exp \left[ \frac{2}{\sigma^2}\left( Dmx + \frac{a-D}{2}x^2) - \frac{1}{4}x^4 \right) \right] \nonumber \\
= & \frac{1}{k+1} \frac{2}{\sigma^2} \times \notag \\
& \int_{-\infty}^{\infty} \diff x \; \left( -(a-D)x^{k+2} + x^{k+4} -Dmx^{k+1} \right) \notag \\
& \times \exp\left[ \frac{2}{\sigma^2} \left( Dmx + \frac{a-D}{2}x^2) - \frac{1}{4}x^4 \right) \right]
\end{align}
for $k \in \mathbb{Z}$, $k\ne -1$. Dividing by the normalization $N_{0}$ we obtain
\begin{align}
(k+1)\langle x^k \rangle = & \frac{2}{\sigma^2} ( -(a-D)\langle x^{k+2} \rangle + \langle x^{k+4} \rangle  \label{eq:recformulae}\\
& -Dm\langle x^{k+1} \rangle ). \nonumber 
\end{align}
In fact Eq.~\eqref{eq:recformulae} is also true for $k=-1$, then $\langle x^3\rangle = a \langle x\rangle$, cf. Eq. \eqref{eq:limR}.

For $a \le a_c$ we have $m =0$, such that all odd moments are zero by symmetry. In that case the recursion formula \eqref{eq:recformulae} simplifies for all even moments to
\begin{align}
	\langle x^{2k+4} \rangle = \frac{\sigma^2}{2}(2k+1) \langle x^{2k} \rangle + (a-D)\langle x^{2k+2} \rangle \label{eq:evenmoms}
\end{align}
with $k=0, 1, \dots$.

At $a=a_c$ we have already calculated $\langle x^2 \rangle|_{a_c} = \sigma^2/(2D)$ in Eq.~\eqref{eq:secondmomentcrit} and know $\langle x^0 \rangle = 1$ since $p_s(x, m)$ is normalized. Therefore it is possible to calculate all even moments using \eqref{eq:evenmoms}. Especially one finds 
\begin{align}
	\langle x^4 \rangle|_{a_c} &= a_c \langle x^2 \rangle|_{a_c} = a_c\frac{\sigma^2}{2D}, \label{eq:xfourac} \\
	\langle x^6 \rangle|_{a_c} &= a_c^2 \frac{\sigma^2}{2D} - a_c \frac{\sigma^2}{2} + \frac{3\sigma^4}{4D}. \label{eq:xsixac}
\end{align}

For small $a-a_c = \varepsilon > 0 $, i.e.\ above but close the critical point, we obtain
\begin{align}
	\langle x^2 \rangle &=  \int_{-\infty}^{\infty} \diff x \, x^2 p_s(x) \notag \\
	&=  \langle x^2 \rangle|_{a_c} + \frac{1}{\sigma^2} \left( \langle x^4 \rangle|_{a_c} - \langle x^2 \rangle^2 |_{a_c}  \right) \varepsilon + \mathcal{O}(\varepsilon^2) \notag \\
&=  \frac{\sigma^2}{2D} + \left( \frac{a_c}{2D} - \frac{\sigma^2}{4D^2} \right)\varepsilon + \mathcal{O}(\varepsilon^2) \label{eq:xsqcrit}
\end{align}
using Eqs. \eqref{eq:secondmomentcrit} and \eqref{eq:xfourac}.

\section{Bounds of $a_c$\label{sec:proofac}}

\indent Since the variance of $x^2$ is larger than zero for any extended distribution we have
\begin{align}
	\langle x^4\rangle - \langle x^2\rangle^2 > 0.
	\label{eq:ineq1}
\end{align}
At $a = a_c$ we obtain with Eqs. \eqref{eq:secondmomentcrit} and \eqref{eq:xfourac}
\begin{align}
	\frac{\sigma^2 a_c}{2D} > & \left( \frac{\sigma^2}{2D} \right)^2, 
\end{align}
which gives the lower bound
\begin{align}
	a_c > & \frac{1}{2} \frac{\sigma^2}{D}.
	\label{eq:ineq2}
\end{align}

To obtain the upper bound we use the inequality
\begin{align}
	\langle x^4\rangle|_{a_c} - 3 \langle x^2\rangle^2 |_{a_c} < 0,	
	\label{eq:kurtosis1}
\end{align}
which states that the kurtosis of $x$ is negative at $a=a_c$, see below. Again with Eqs. \eqref{eq:secondmomentcrit} and \eqref{eq:xfourac} we find
\begin{align}
	\frac{\sigma^2 a_c}{2D} < & 3\left( \frac{\sigma^2}{2D} \right)^2,
\end{align}
which gives
\begin{align}
	a_c < & \frac{3}{2} \frac{\sigma^2}{D}.
	\label{eq:acineq2}
\end{align}

To show \eqref{eq:kurtosis1} we substitute
\begin{align}
	x =  \sqrt{\langle x^2 \rangle} \, y 
	\label{eq:subst}
\end{align}
such that the new coordinate $y$ has variance one. We denote the stationary distribution of the new coordinate by $p(y)$. The inequality~\eqref{eq:kurtosis1} in the $x$ coordinate is equivalent to the same expression in the new coordinate $y$
\begin{align}
	\langle y^4\rangle -3\langle y^2\rangle^2 < 0.
	\label{eq:kurtnewcoord}
\end{align}

Now we compare the distribution $p(y)$ with the Gaussian distribution with variance one denoted by $g(y)$. In the following we only consider the critical point $a=a_c$ where both distributions have zero mean and are symmetric under the transformation $y \rightarrow -y$. We look at the intersection of both curves and distinguish two cases. There can be either two or four intersecting points.

\indent For two intersection points we use the theorem \cite{MJ99}:

\noindent If a symmetric zero-mean probability distribution $p(y)$ intersects with the standard normal distribution $g(y)$ in exactly two points $-y_0, y_0$, then $g(y) > p(y)$ for all $y>y_0$ if and only if the kurtosis of $p(y)$ is negative.

\noindent In the present situation $p(y)$ decays as $\exp(- \lambda y^4)$, \mbox{$\lambda>0$} as $y \rightarrow \pm \infty$ and therefore $g(y) > p(y)$ for large enough $|y|$. Hence we can apply the theorem and \eqref{eq:kurtnewcoord} is satisfied.

In the case of four intersection points of $p(y)$ and $g(y)$ we use the following theorem from \cite{Dy43}:

\noindent Suppose two probability densities $g(y)$ and $p(y)$ with zero mean and the same variance are given. Let $\mu_{g3}, \mu_{g4}; \, \mu_{p3}, \mu_{p4}$ be their respective third and fourth moments. Then we have a sufficient condition for \mbox{$\mu_{g4}\ge \mu_{p4}$}: there should exist four abscissae $a_1<a_2<a_3<a_4$ such that\\
\begin{align}
	\text{(i) when} \quad  \begin{array}{r}-\infty < y < a_1 \\
	a_2 < y < a_3  \\
	a_4 < y < \infty \\ \end{array} \Bigg\} \, \, \, g(y) \ge p(y),
\end{align}
\begin{align}
	\text{and (ii) when} \quad  \begin{array}{r}
	a_1 < y < a_2 \\
a_3 < y < a_4 \\  \end{array} \Big\} \, \, \, g(y)\le p(y),
\end{align}
and (iii) $a_1+a_2+a_3+a_4$ and $\mu_{p3} - \mu_{g3}$ are not both strictly positive or both strictly negative.

\noindent In the present case we have $\mu_{g3} = \mu_{p3} = 0$. Furthermore $a_1 = -a_4$ and $a_2 = - a_3$ since both $g(y)$ and $p(y)$ are even functions. Hence $a_1 + a_2 + a_3 + a_4 =0$ and we can apply the theorem. Therefore $\mu_{p4} \le \mu_{g4}$ \footnote{Dawson \cite{Da83} already showed that the fourth cumulant is less or equal to zero, cf. Eq. $(4.15)$ there.  By numerical evidence he conjectured that there is strict inequality. We prove this conjecture.}. We can follow the lines in \cite{Dy43} to prove even strict inequality.

\noindent Consider the function $h(y) = (a_1-y)(a_2-y)(a_3-y)(a_4-y)$. For any $y \in \mathbb{R}$ the functions $g(y)-p(y)$ and $h(y)$ have either the same sign or at least one of them is zero. Thus $h(y)\left(g(y) - p(y) \right) \ge 0$. In the present situation, since both functions are continuous and zero only if $y\in \{a_1, a_2, a_3, a_4 \}$, there exists $\varepsilon, \delta >0$ such that $h(y)\left(g(y) - p(y) \right)>\varepsilon$ for $y\in \mathbb{R}$, $y \not\in [a_i- \delta, a_i + \delta]$ for $i=1, 2, 3, 4$. Hence we have
\begin{align}
	\int_{-\infty}^{\infty} \diff y \, h(y)\left(g(y) - p(y) \right) > 0.
	\label{eq:strictinequ}
\end{align}
Expanding the polynomial $h(y)$ and performing the integral in \eqref{eq:strictinequ} we find
\begin{align}
	\mu_{g4} -\mu_{p4} >0,
	\label{eq:fourthmomentiequ}
\end{align}
where we used that odd moments of $p(y)$ and $g(y)$ are zero and that both distributions have variance one and are normalized. Since the kurtosis of a Gaussian is zero, by \eqref{eq:fourthmomentiequ} we follow that \eqref{eq:kurtnewcoord} is true, which completes the proof of the inequality \eqref{eq:ineqac}.

\section{Laplace Approximation\label{sec:laplace}}
In the following we solve the PTC \eqref{eq:ptcphi} asymptotically for strong and for weak noise using Laplace's method to evaluate the integrals in Eq. \eqref{eq:varphi}. Independently of the simpler method using the ODE \eqref{eq:ode} we obtain the same results. When using Laplace's method we explicitly use the different shapes of the distribution $p_s(x)$ for weak and strong noise, discussed in section \ref{subsec:scaling}. In the next section we explicitly calculate the Integral $\int_{-\infty}^{\infty} \diff x \exp(\lambda \Phi(x))$ for large $\lambda$, where $\Phi(x)$ is a polynomial of degree four. In the succeeding two sections we apply the results of this calculation to asymptotically solve the PTC \eqref{eq:ptcphi} for weak and for strong noise.
\subsection{Evaluation of the Integral}
\noindent We evaluate an integral of the form
\begin{align}
	I =  \int_{-\infty}^{\infty} \diff x \exp (\lambda \Phi(x))
	\label{eq:intlp1}
\end{align}
for large $\lambda$ applying the Laplace method \cite{BO78}. We denote
\begin{align}
	\frac{d^n}{dx^n}\Phi(0) :=  \Phi^{(n)}_{0}.
	\label{eq:abbrder}
\end{align}
Suppose that $\Phi(x)$ has its global maximum at $x=0$. Expanding around zero yields
\begin{align}
	\Phi(x) =  \Phi_0 + \frac{1}{2}\Phi^{(2)}_0 x^2 + \frac{1}{6}\Phi^{(3)}_0 x^3 + \frac{1}{24}\Phi^{(4)}_0 x^4 + \cdots.
	\label{eq:expandlp1}
\end{align}
Now we change the integration boundaries in \eqref{eq:intlp1} to $-\varepsilon$ and $\varepsilon$ for some $\varepsilon>0$. Doing so, we make only exponentially small errors for $\lambda \rightarrow \infty$,
\begin{align}
	I \approx  \int_{-\varepsilon}^{\varepsilon}dx \exp(\lambda \Phi(x)).
	\label{eq:intlp2}
\end{align}
Inserting the expansion \eqref{eq:expandlp1} yields
\begin{align}
	I \approx & \exp(\lambda \Phi_0 ) \notag \\ 
	& \times \int_{-\varepsilon}^{\varepsilon} \diff x \exp\left[ \lambda \left( \frac{1}{6} \Phi^{(3)}_0 x^3 + \frac{1}{24} \Phi^{(4)}_0 x^4 + \cdots \right) \right] \notag \\
	& \times \exp\left(\lambda \frac{1}{2} \Phi^{(2)}_0 x^2\right).
	\label{eq:intlp3}
\end{align}
We are interested in the case where $\Phi$ is a polynomial of fourth order. We expand $g(x):=\exp(\lambda ( \frac{1}{6} \Phi^{(3)}_0 x^3 + \frac{1}{24} \Phi^{(4)}_0 x^4))$ for small $x$ and neglect terms which give either a vanishing contribution or contributions of the order $\lambda^{-9/2}$ when the Gaussian integrals are performed. In view of
\begin{align}
	\int_{-\infty}^{\infty} \diff x \, x^{2n} & \exp\left[ -\alpha x^2 \right] =  \frac{(2n-1)!!}{2^n}\frac{1}{\alpha^n}\sqrt{\frac{\pi}{\alpha}}
	\label{eq:gauss4},\\ 
	&\int_{-\infty}^{\infty} \diff x \, x^{2n+1}\exp\left[ -\alpha x^2 \right] =  0 
\end{align}
we should take into account \cite{mathematica}
\begin{align}
	g(x) = & \, 1 + \frac{1}{24}\lambda \Phi^{(4)}_0 x^4 + \frac{1}{72}\lambda^2 {\Phi^{(3)}_{0}}^{2} x^6  \notag \\
	& + \frac{1}{1152}\lambda^2 {\Phi^{(4)}_0}^2 x^8 +\frac{1}{1728} \lambda^3 \Phi^{(4)}_0 {\Phi^{(3)}_0}^2 x^{10} \notag \\
       & + \frac{1}{248832}(8\lambda^4 {\Phi^{(3)}_0}^4 + 3 \lambda^3 {\Phi^{(4)}_0}^3)x^{12} \notag \\
       & + \frac{1}{82944} \lambda^4 {\Phi^{(4)}_0}^2 {\Phi^{(3)}_0}^2 x^{14} \notag \\
       & + \frac{1}{746496} \lambda^5 \Phi^{(4)}_0 {\Phi^{(3)}_0}^4 x^{16} \notag \\
       & + \frac{1}{33592320} \lambda^6 {\Phi^{(3)}_0}^6 x^{18} + \cdots
	\label{eq:expandlp2}
\end{align}
Now we plug the expansion \eqref{eq:expandlp2} into Eq.~\eqref{eq:intlp3} and extend the integration range to $(-\infty, \infty)$. Again the change of the integration range produces only an exponentially small error for $\lambda\rightarrow \infty$. Evaluating the Gaussian integrals leads to
\begin{align}
	I = & \exp(\lambda \Phi_0) \sqrt{\frac{2\pi}{-\lambda \Phi^{(2)}_0}} \label{eq:intlp4}\\
	&\times \Bigg[ 1 + \lambda^{-1}\left(\frac{1}{8} \frac{\Phi^{(4)}_0}{ {\Phi^{(2)}_0}^2} - \frac{5}{24} \frac{ {\Phi^{(3)}_0}^2}{ {\Phi^{(2)}_0}^3}\right) \notag \\
		& + \lambda^{-2} \left( \frac{35}{384} \frac{ {\Phi^{(4)}_0}^2}{ {\Phi^{(2)}_0}^4} - \frac{35}{64} \frac{\Phi^{(4)}_0 {\Phi^{(3)}_0}^2}{ {\Phi^{(2)}_0}^5} + \frac{385}{1152} \frac{ {\Phi^{(3)}_0}^4}{ {\Phi^{(2)}_0}^6}\right) \notag \\
		& + \lambda^{-3} \Bigg( \frac{385}{3072} \frac{ {\Phi^{(4)}_0}^3}{ {\Phi^{(2)}_0}^6} - \frac{5005}{3072}\frac{ {\Phi^{(4)}_0}^2 {\Phi^{(3)}_0}^2}{ {\Phi^{(2)}_0}^7} \notag \\
		& + \frac{25025}{9216}\frac{\Phi^{4}_0 {\Phi^{(3)}_0}^4}{ {\Phi^{(2)}_0}^8} - \frac{85085}{82944} \frac{ {\Phi^{(3)}_0}^6}{ {\Phi^{(2)}_0}^9} \Bigg) +  \mathcal{O}(\lambda^{-4}) \Bigg]. \notag
\end{align}

\subsection{Weak Noise\label{sec:laplacewn}}

\noindent In the limit of weak noise we set $\lambda = 1/\sigma^{2}$ and use the ansatz
\begin{align}
	a_c(D, \sigma)= \sum_{n=1}^{\infty}a_{n}(D)\sigma^{2n},
	\label{eq:ansatzwnapp}
\end{align}
which is equivalent to Eq. \eqref{eq:ansatzwn}. We define 
\begin{align}
	\Phi(x) := \left(a_c(\sigma, D) - D\right)x^2 - \frac{x^4}{2}, \label{eq:phix} \\
I := \int_{-\infty}^{\infty} \diff x \, \exp\left[ \frac{1}{\sigma^2} \Phi(x) \right] .
\label{eq:integr1}
\end{align}
and rewrite the PTC \eqref{eq:ptcphi} as
\begin{align}
	1 = \frac{2D}{\sigma^2} \partial_{a_1}\ln I \label{eq:ptcweaknoise}
\end{align}
to express the right hand side of Eq.~\eqref{eq:ptcphi} in a controllable way as a series in $\sigma^2$.\\
In view of the limit \eqref{eq:a_cconvwn} it is clear that $\Phi(x; \sigma)$ is locally uniformly converging to
\begin{align}
\Phi_0(x) := -D x^2 - \frac{x^4}{2}
\end{align}
as $\sigma \rightarrow 0$ such that we can use Laplace's method \cite{BO78} to evaluate the integral \eqref{eq:integr1}. For $\Phi(x)$ defined by \eqref{eq:phix} we find %
$\Phi_0=\Phi_0^{(1)}=\Phi_{0}^{(3)} = 0$ and
\begin{align}
	\Phi_{0}^{(2)} =  2(a_c - D), \quad \Phi_{0}^{(4)} =  -12 \label{eq:Phiwn5}.
\end{align}
Inserting Eq. \eqref{eq:intlp4} into the PTC \eqref{eq:ptcweaknoise} leads to
\begin{align}
	1 = & \, 2D \Bigg[ -\frac{1}{\Phi_{0}^{(2)}} + \Bigg( -\sigma^{2}\frac{1}{2}\frac{\Phi_{0}^{(4)}}{ {\Phi_{0}^{(2)}}^3} -\sigma^{4}\frac{35}{48} \frac{ {\Phi_{0}^{(4)}}^2}{ {\Phi_{0}^{(2)}}^5}  \label{eq:PTC}\\
	      &	- \sigma^{6}\frac{385}{256} \frac{ {\Phi_{0}^{(4)}}^3}{  {\Phi_{0}^{(2)} }^7} + \mathcal{O}(\sigma^{8}) \Bigg)\Big/ \Bigg( 1 + \sigma^{2}\left(\frac{1}{8} \frac{ \Phi_{0}^{(4)}}{ {\Phi_{0}^{(2)}}^2} \right) \notag \\
		& + \sigma^{4} \left( \frac{35}{384} \frac{ {\Phi_{0}^{(4)}}^2}{ {\Phi_{0}^{(2)}}^4} \right)  + \sigma^{6}\frac{385}{3072} \frac{ {\Phi_{0}^{(4)}}^3}{ {\Phi_{0}^{(2)}}^6} + \mathcal{O}(\sigma^{8}) \Bigg) \Bigg].
	\notag
\end{align}
Inserting the ansatz \eqref{eq:ansatzwnapp} into Eq. \eqref{eq:PTC} and expanding in powers of $\sigma^{2}$
leads to \cite{mathematica}
\begin{align}
	1 = & \, 1 + \frac{2a_1D-3}{2D^2}\sigma^{2} + \notag \\
	& + \frac{2a_2D^3+2a_1^2D^2-9a_1D+12}{2D^4}\sigma^{4} + \notag \\
	& + \frac{1}{8D^6} \Bigg( 8a_3D^5 + 16a_1a_2D^4 +\left(-32 a_2 + 8a_1^3 \right)D^3 \notag \\
       & - 72 a_1^2 D^2 + 240 a_1 D - 297 \Bigg) \sigma^{6} + \mathcal{O}(\sigma^{8}).
	\label{eq:PTC2}
\end{align}
Comparing coefficients in powers of $\sigma^2$, we find
\begin{align}
	a_1 =  \frac{3}{2D},\quad 
	a_2 =  -\frac{3}{2D^3},\quad
	a_3 =  \frac{27}{4D^5},
	\label{eq:coefficients}
\end{align}
which is equivalent to the result \eqref{eq:weaknoiseac2}.

\subsection{Strong Noise\label{sec:laplacesn}}
\noindent In the limit of strong noise we set $\lambda = \sigma^2$ and use the ansatz
\begin{align}
	a_c(D, \sigma) = a_1(D) \sigma^2 + \sum_{n=0}^{\infty} a_{-n}(D)\beta^{-n},
	\label{eq:ansatzsnapp}
\end{align}
which is equivalent to Eq. \eqref{eq:ansatzsn}. We substitute \mbox{$x=\sigma\cdot y$} and define
\begin{align}
	\tilde{I} = \frac{1}{\sigma} I =  \int_{-\infty}^{\infty} \diff y \, \exp\left[\sigma^2 \Psi(y) \right]  \label{eq:intstrongnoise}
\end{align}
with the function
\begin{align}
	\Psi(y) =  \frac{a_c(\sigma, D)-D}{\sigma^2}y^2 - \frac{y^4}{2} , \label{eq:psiy}
\end{align}
that has its local maxima at
\begin{align}
	y_{\text{max}} = \pm \frac{\sqrt{a_c-D}}{\sigma}.
	\label{eq:ymax}
\end{align}
By the inequality \eqref{eq:ineqac} we find that $a_0 > 0$. Analogously to the weak noise case we rewrite the PTC \eqref{eq:ptcphi} as
 \begin{align}
	 1 =  \frac{2D}{\sigma^2} \partial_{a_1}\ln \tilde{I}. \label{eq:ptcstrongnoise}
 \end{align}
For strong noise, $\Psi(y)$ is locally uniformly converging to
\begin{align}
\Psi_0(y) = a_1 y^2 - \frac{y^4}{2}.
\end{align}
We will perform Laplace approximation around the positive maximum, an equal contribution comes from the other maximum. Instead of $\Phi_{0}^{(n)}$ in \eqref{eq:intlp4} we use $\Psi^{(n)}(y_\text{max})$ with
\begin{align}
	\Psi(y_\text{max}) &=  \frac{(a_c-D)^2}{2\sigma^4}, \\
	\Psi^{(1)}(y_{\text{max}}) &=  0, \\
	\Psi^{(2)}(y_{\text{max}}) &=  -4 \frac{a_c-D}{\sigma^2}, \\
	\Psi^{(3)}(y_{\text{max}}) &=  - 12 \frac{\sqrt{a_c-D}}{\sigma} ,\\
	\Psi^{(4)}(y_{\text{max}}) &=  - 12 .
	\label{eq:derpsisn}
\end{align}
By \eqref{eq:intlp4} with $\tilde{I}$ instead of $I$ we obtain up to $\mathcal{O}(\sigma^{-6})$
\begin{align}
	\ln \tilde{I} = & \, \ln 2 + \sigma^2 \frac{\tilde{a}^2}{2} + \frac{1}{2} \ln (2\pi) - \frac{1}{2} \ln (\sigma^{2}4\tilde{a}) \notag \\
	& + \ln \Bigg[ 1 + \sigma^{-2} \left( - \frac{3}{32} \frac{1}{\tilde{a}^2} + \frac{15}{32} \frac{1}{\tilde{a}^2}  \right)  \notag \\
		& + \sigma^{-4} \Bigg( \frac{105}{2048} \frac{1}{\tilde{a}^4} - \frac{945}{1024} \frac{1}{\tilde{a}^4} + \frac{3465}{2048} \frac{1}{\tilde{a}^4}  \Bigg) \Bigg] \notag \\
	= & \,  \sigma^2 \frac{\tilde{a}^2}{2} + \frac{1}{2} \ln (2\pi) - \frac{1}{2}\ln (\sigma^{2}\tilde{a})  \notag \\
	& + \ln \Bigg[ 1 + \sigma^{-2} \frac{3}{8} \frac{1}{\tilde{a}^2} +\sigma^{-4} \frac{105}{128} \frac{1}{\tilde{a}^4} \Bigg],
	\label{eq:lnistrnoise}
\end{align}
with 
\begin{align}
	\tilde{a} = \frac{a_c -D}{\sigma^{2}}.
	\label{eq:atilde}
\end{align}
Inserting \eqref{eq:lnistrnoise} into the PTC~\eqref{eq:ptcstrongnoise} we obtain
\begin{align}
	1=& \frac{2D}{\sigma^{2}} \Bigg[\sigma^{2} \tilde{a} - \frac{1}{2\tilde{a}} \label{eq:ptcsn2} - \left( \sigma^{-2} \frac{3}{4} \frac{1}{\tilde{a}^3} + \sigma^{-4} \frac{105}{32} \frac{1}{\tilde{a}^5} \right) \Big/  \\
		& \Bigg(  1 + \sigma^{-2} \frac{3}{8} \frac{1}{\tilde{a}^2} + \sigma^{-4} \frac{105}{128} \frac{1}{\tilde{a}^4}  \Bigg)\Bigg]. \notag
\end{align}
Inserting \eqref{eq:atilde} and furthermore the ansatz for $a_c$ Eq. \eqref{eq:ansatzsnapp} we expand in powers of $\sigma^{-2}$ and obtain \cite{mathematica}
\begin{align}
	1 = & \, 2a_1D + \sigma^{-2}\frac{2a_1a_0 D - D - 2a_1D^2}{a_1} \notag \\
	&+ \sigma^{-4}\frac{2 a_1 a_0 D + 4 a_1^3a_{-1}D - 3D - 2 a_{1}D^2 }{2a_1^3}.
	\label{eq:ptcsn3}
\end{align}
Comparing coefficients in powers of $\sigma^{2}$ leads to
\begin{align}
	a_1 =  \frac{1}{2D}, \quad a_0 =  2D, \quad a_{-1} =  4 D^3, 
	\label{eq:resultsn}
\end{align}
which is equivalent to Eq.~\eqref{eq:strongnoiseac}.


%

\end{document}